\documentclass[]{article}
\usepackage{multicol}
\usepackage{apj}
\newcommand {\h} {$h^{-1} \, Mpc^{-1} \,$}
\newcommand {\ks} {$km~s^{-1} \;$}

\newcommand {\msun} {$h^{-1} \  M_{\odot} \;$}
\newcommand {\m} {$M_{\odot} \;$}
\newcommand {\arcmin} {\hbox{$^\prime$}}
\newcommand {\arcsec} {\hbox{$^{\prime\prime}$}}
\begin{document}

\vspace{15mm}
\begin{center}
\uppercase{
New Optical Insights into the Mass Discrepancy of Galaxy
Clusters:\\
The  Cases of A1689 and A2218}\\
\vspace*{1.5ex}
{\sc 
M. Girardi$^{1,2,3}$, D. Fadda$^{1,2,3}$, E. Escalera$^{2,3,4}$,
 G.  Giuricin$^{2,3}$, F. Mardirossian$^{1,2,3}$ and
M. Mezzetti $^{2,3}$}\\
%\vspace*{-1pt}
\vspace*{1.ex}
{\small 
$^1$Osservatorio Astronomico di Trieste, Italy\\ 
$^2$SISSA, via Beirut 4, 34013 - Trieste, Italy\\ 
$^3$Dipartimento di Astronomia, Universit\`{a} degli Studi di Trieste, \\ 
$^4$Observatoire de Marseille, Place Le Verrier, F-13248, 
Marseille, C\'edex 4, France \\ 
E-mail: girardi, fadda, giuricin, mardiros, mezzetti @sissa.it}
\end{center}
\vspace*{-6pt}

\begin{abstract}
The discrepancy between the masses estimated from X-ray and from
strong gravitational lensing analyses was recently pointed out for
several clusters and in particular for A1689 and A2218.

We analyze the internal structures of these clusters by applying a
recent development of the method of wavelet analysis, which uses the
complete information obtained from optical data, i.e. galaxy positions
and redshifts.  We find that both clusters show the presence of
structures superimposed along the line of sight with different mean
redshifts and smaller velocity dispersions than that of the system as
a whole, suggesting that the clusters could be cases of the on-going
merging of clumps.

We compute the masses of the
two clusters by adding the optical virial masses of the single clumps, which
are supposed to be virialized. 
By rescaling our masses to the same radii, we compare our
results with mass estimates derived from X-ray and 
gravitational lensing analyses.   

In the case of A2218 we find an acceptable agreement with X-ray
and  gravitational lensing masses.

On the contrary, in the case of A1689 we find that our mass estimates
are smaller than X-ray and gravitational lensing ones at both small
and large radii.  But the observed X-ray temperature in this cluster,
and thus the derived X-ray mass, could be enhanced by collision-heated
gas during cluster merging. The high central cluster mass obtained
from lensing analysis could be explained only by adding the masses of
two background galaxy groups possibly aligned with the cluster.

In any case, at variance with earlier claims, there is no
evidence that X-ray mass estimates are underestimated.\\
\vspace*{-6pt}
\noindent
{\em Subject headings: } 
galaxies: clusters: individual (A1689, A2218) - cosmology:
observations

\end{abstract}  
\begin{multicols}{2}

%%%%%%%%%%
\section{INTRODUCTION}

The recent literature points out that present estimates of cluster
masses derived from X-ray data analysis are smaller than the respective
estimates derived from gravitational lensing analysis (e.g. Wu \& Fang
1997, hereafter WF97).  This discrepancy is especially evident in the
case of strong gravitational lensing (see Fig.~2 of Wu \& Fang 1996).
Indeed, some recent papers outline the presence of this discrepancy in
a few galaxy clusters, which show long arcs and arclets produced by
gravitational lensing, and for which the radial density distribution
and temperature of the intracluster gas are known (e.g. Wu 1994;
Miralda-Escud\'e \& Babul 1995, hereafter MB95).

Apart from the intrinsic interest of knowing the masses of galaxy
clusters, understanding the mass discrepancy is also important for
cosmological reasons.  If cluster masses provided by X-ray analysis
are really underestimated, the cluster baryon fraction is reduced and
one can solve, at least partially, the recently claimed baryon crisis
in clusters of galaxies and thus reconcile observations with the
value of $\Omega=1$ (e.g. White et al. 1993; White \& Fabian 1995).

The general reliability of cluster mass estimates derived from X-ray
analysis, as checked by numerical simulations, is still a matter of
discussion (e.g. Schindler 1996; Evrard, Metzler, \& Navarro 1996; see
Bartelmann \& Steinmetz 1996 regarding a systematic mass underestimation).
However, the authors agree that, when a strong deviation from hydrostatic
equilibrium occurs, the X-ray techniques can strongly overestimate or
underestimate the mass depending on, e.g., the presence of shock
waves, subclustering and the phase of the merging
event, with a tendency to underestimate the total mass throughout the
post-merger phase (Roettiger, Burns, \& Loken 1996). This last fact,
coupled with the suggestion that the probability for a cluster to form
large arcs is significantly higher than the norm 
if the cluster is substructured
(Miralda-Escud\'e 1993; Bartelmann, Steinmetz, \& Weiss 1995), could explain the mass
discrepancy.

The reliability of gravitational lensing masses also seems to be
supported by the agreement between the mass estimated by means of
optical and gravitational lensing methods as found in the pioneering
work by WF97, who used a sample of 29 lensing clusters.  However, they
used published velocity dispersions and an isothermal sphere model in
order to estimate optical virial masses, without taking into account
the internal structures of clusters.  Indeed, estimates of velocity
dispersions and optical masses for bimodal or complex clusters
strongly depend on whether they are treated as single systems or as
sums of different clumps (see Girardi et al. 1997, hereafter G97 and
references therein).

%%FIGURE 1%%%
\end{multicols}
\includegraphics{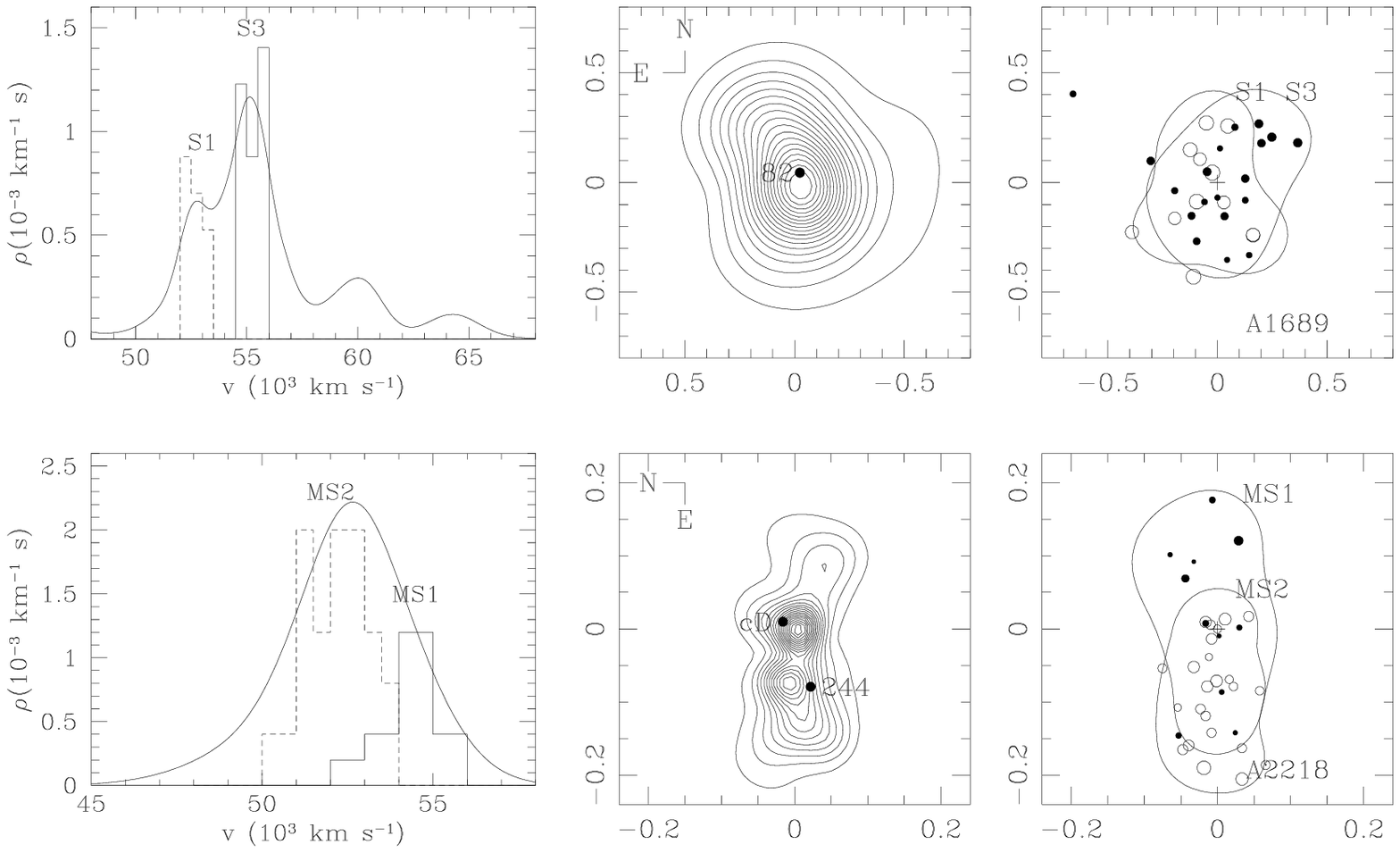}
$\ \ \ \ \ \ $\\
\vspace{11.4cm}
$\ \ \ $\\
%\vspace{-20mm}
{\small\parindent=3.5mm
{\sc Fig.}~1.---
From left to right: the velocity distribution,
the density map, and the map with highlighted structures for clusters
A1689 (at top) and A2218 (at bottom). The higher the redshifts of
galaxies, the smaller the symbol sizes. Distances from the density
center are in Mpc.  The velocity distributions of A1689 and A2218 show
(in the form of histograms) the contributes of the two main
substructures to the total distributions.  The velocity distribution
of A1689 also shows the two background structures.
}
\vspace{5mm}
\begin{multicols}{2}

  Recent improvements in three-dimensional
structure analysis and its wide application to nearby clusters (Serna
\& Gerbal 1996; G97) suggest the need to reanalyze the question of optical
masses.

Here we analyze clusters A1689 and A2218, which have a
large enough number of galaxies with published redshifts (131 and 54,
respectively).

MB95 analyzed three clusters at a
redshift of 0.17-0.20 and used long arcs to compute
the projected mass within the inner core which is then deprojected
and extrapolated to be compared to the X-ray mass. They
found that the mass derived from strong lensing
is larger by a factor of 2-2.5 than the mass estimated from the X-ray
emission of the intracluster gas in clusters A1689 and A2218. The
results are more ambiguous for the third cluster A2163.
The explanation of the
discrepancy could reside, e.g., in non-isothermality, inhomogeneity,
and the presence of non-thermal pressure support of the gas, as well
as in the highly prolate structure of the cluster potential, i.e. of
the lens. According to MB95, none of these explanations seems to
account for the entire problem. In particular, for clusters which are
reasonably elongated along the line of sight, the mass evaluated by
means of strong gravitational lensing can be overestimated by only a
factor of 1.6-2.0 (Bartelmann 1995). However, in the case of A2218, 
recent work by Squires
et al. (1996) shows a good agreement between X-ray and weak
gravitational mass at 0.4 \h. The optical masses of A1689 and A2218,
as estimated by Wu \& Fang (1997) at several apertures, are generally
larger than both respective X-ray and gravitational lensing estimates.

In \S~2 we present our structure analysis, on which we base our
estimates of optical virial masses of the two clusters (\S~3). 
In \S~4 we compare our optical mass estimates with those derived from
X-ray and gravitational lensing analyses. 
In \S~5
we outline our discussions, and in \S~6 we summarize our findings and
conclusions.

Throughout the paper we give errors at the 68\% confidence level
(hereafter c.l.)  and we use H$_0$ = 100 $h$ $Mpc^{-1}$ \ks.

\section{ANALYSIS OF CLUSTER STRUCTURES}

The redshift data of A1689 galaxies come from Teague, Carter, \& Gray
(1990) and the magnitude data, only partially available,  from
Gudehus \& Hegyi (1991). 

%%FIGURE 2%%%
\end{multicols}
\includegraphics{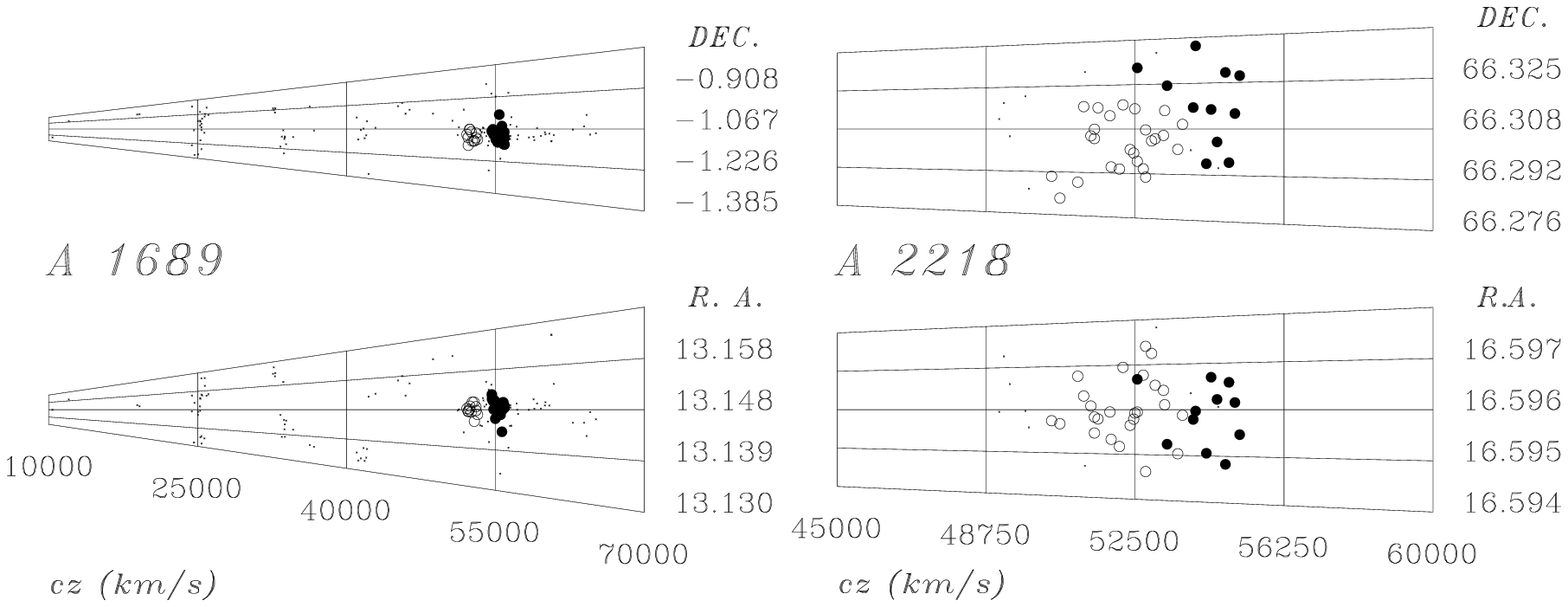}
$\ \ \ \ \ \ $\\
\vspace{8.4cm}
$\ \ \ $\\
%\vspace{-20mm}
{\small\parindent=3.5mm
{\sc Fig.}~2.---
The wedge diagrams  for A1689 and A2218. All galaxies
with redshift in the original samples are shown. The main substructures
are evidenced: S1 and S3 (open and solid circles, respectively) for A1689
and MS1 and MS2 (solid and open circles, respectively) for A2218.}
\vspace{5mm}
\begin{multicols}{2}

The galaxy data for A2218 come from Le
Borgne, Pell\'o, \& Sanahuja (1992); in this case only the central
cluster region is sampled.  The samples we use are suitable for
searching for substructures.  In fact, the A1689 data consist in a
random sample extracted from a complete magnitude sample. Moreover,
the A2218 sample shows no correlation between galaxy magnitude and the
distance from cluster center. In this way, although a
magnitude-complete sample would be preferable, we avoid having some
oversampled cluster regions which produce spurious substructures.

Figure~1 shows the velocity distributions and the galaxy density maps
of A1689 and A2218 as obtained according to the one- and
two-dimensional adaptive kernel methods (e.g. Pisani 1993;
see also Girardi et al. 1996 for a wide application to nearby clusters).

  However, our structural analysis is based on the recent procedure of
structure analysis by G97 which uses combined galaxy and redshift
data.
This method is an improvement over that described in Escalera
\& Mazure (1992), and couples the technique of wavelet analysis with
local kinematical estimators, which are computed for each galaxy.  The
technique has the advantage of assigning, within a certain range of
probability, the galaxies to the detected structure, thus enabling us
to make kinematical and dynamical analyses of the cluster. We
refer to G97 for a more detailed description of the method.

First, we roughly extract the main peak (MP) in the velocity
distribution. Subsequently, we perform three-dimensional analysis.
The structure detected at the largest scale (with respect to the whole
sampled region) is the main system MS, or MS1 and MS2 in the
case of a bimodal structure.  At smaller scales the method uncovers the
presence of internal substructures (S).  All the structures hereafter
mentioned have a confidence level greater than 99.5\%; i.e. there are
fewer than 5 chances out of 1000 due to random galaxy association.
The statistical significances are derived by comparing
the wavelet coefficients obtained in the real field to those
produced in a series of N replicas which are obtained by drawing independently
the positions $X_i$ and $Y_i$ from the $X$ and $Y$ distributions of the
sample studied and then by randomly reassigning the velocities (see
e.g. Escalera \& Mazure 1992).

\noindent {\sl A1689} --- By using the one-dimensional adaptive kernel method, 
Girardi et al. (1996) found that cluster A1689 consists of two
strongly overlapping peaks in the velocity distribution (see
Figure~1).  In particular, the method assigns 16 and 41 galaxies to
the two peaks, but 15 galaxies assigned to one peak have a high
probability of belonging to the other peak. In Fig.~1, which shows only
the range of radial velocities between $\sim 50000-70000$ \ks of the
whole field, also two background (not significant) peaks are shown.
Instead, A1689 appears to be regular in the galaxy density map.  

By applying the multi-scale analysis of G97, 
A1689 appears to be a condensed structure including three distinct groups,
which fully overlap with one another, but are extremely well separated in
terms of velocity: $\Delta$ $V_{1-2} \sim 1200$ \ks and $\Delta$
$V_{2-3} \sim$ 1500 \ks.  The first group (S1) is a coherent
structure, compact and regular. On the contrary, the second one (S2)
is a collection of intermediate galaxies, namely three pairs and a
loose object.  Actually S2 presents only two of its seven galaxies
within the supposed virialization radius and  so we do not
consider S2 in the mass estimate of A1689 (see \S~3).  The third one (S3)
is the most extended and populated; thus it appears to be the dominant
system. Since substructures represent most of the galaxy population,
we define this cluster as {\em complex} according to the G97
morphological classification.\\

%%TAB1
\vspace{4mm}
\hspace{-4mm}
\begin{minipage}{9cm}
\renewcommand{\arraystretch}{1.2}
\renewcommand{\tabcolsep}{1.2mm}
\begin{center}
\vspace{-3mm}
TABLE 1\\
\vspace{2mm}
{\sc Detected Structures\\}
\footnotesize
\vspace{2mm}
\begin{tabular}{lrrrrr}
\hline \hline
\multicolumn{1}{c}{Name}
&\multicolumn{1}{c}{$N$}
&\multicolumn{1}{c}{Center (1950)}
&\multicolumn{1}{c}{$\overline{V}$}
&\multicolumn{1}{c}{$R$}
&\multicolumn{1}{c}{$\sigma$}
\\
&
&\multicolumn{1}{c}{($\alpha$,$\delta$)}
&\multicolumn{1}{c}{(km/s)}
&\multicolumn{1}{c}{(Mpc)}
&\multicolumn{1}{c}{(km/s)}\\
\hline
\multicolumn{6}{c}{A1689}\\
\hline
MS=MP & 57& 130855.9$-$010504& 54477& 2.26&       1429$_{-96}^{+145}$\\
S1    & 12& 130855.0$-$010442& 52638& 0.38& 321$_{-29\phn}^{+57\phn}$\\
S2    &  7& 130839.6$-$010320& 53827& 0.62& 243$_{-28\phn}^{+84\phn}$\\
S3    & 20& 130856.8$-$010526& 55300& 0.87& 390$_{-27\phn}^{+52\phn}$\\
\hline
\multicolumn{6}{c}{A2218}\\
\hline
MP & 48& 163541.0$+$661854& 52549& 0.23&   1405$_{-129}^{+163}$\\
MS1& 11& 163547.4$+$661845& 54397& 0.17&    660$_{-121}^{+292}$\\
MS2& 25& 163543.0$+$661905& 52229& 0.16& 788$_{-68\phn}^{+128}$\\
S  &  5& 163541.8$+$661900& 51672& 0.06& 449$_{-16\phn}^{+288}$\\
\hline
\end{tabular}

\end{center}
\vspace{3mm}
\end{minipage}
\normalsize

In the lensing model of A1689, MB95 assumed the existence of two
clumps, which correspond to a large central concentration around
galaxy no. 82 in Gudehus \& Hegyi (1991) and to a smaller
concentration of galaxies lying 
$1\arcmin$ to the north-east.  Also the centers of
our structures S1 and S3 are aligned along the north-east direction,
but they do not correspond to those of MB95.  The center of S1 roughly
coincides with galaxy no. 82, though the galaxy itself belongs to
S3.  The center of S3 lies about $50\arcsec$
 to the south-west of S1.

\noindent {\sl A2218} --- The one-dimensional adaptive method applied
to the cluster A2218 gives a velocity distribution with a single peak,
slightly asymmetrical (skewness = $-$ 0.44 $\pm$ 0.35), while the
cluster shows two concentrations in the galaxy density map (see Figure~1).

The redshift information on A2218 allows us to study only the central
cluster region, which consists of two main structures (MS1 and MS2)
separated by about 2000 \ks and superimposed along the line of
sight. A further small substructure (S) is detected in the more
populated MS2 structure and contains the cD galaxy.  
We define this cluster as {\em bimodal} according to the G97
morphological classification.

In the lensing model of A2218, MB95 (see also Kneib et al. 1995)
assumed as principal clumps those centered on the cD galaxy and on 
galaxy no. 244 in Le Borgn\'e et al. (1992), which roughly correspond
to the two most significant clumps in our two-dimensional galaxy
distribution (Figure~1).  Also our structures are aligned along the
east-west direction, but the more heavily populated one, MS2, lies eastward
and contains both the cD galaxy and galaxy no. 244. \\

Table~1 presents the results from the multi-scale structure analysis
and related kinematical analysis (the structure type is
indicated by the symbols described above). For each structure we list the
number $N_{gal}$ of galaxies involved; the coordinates 
of the galaxy density center computed by using the
two-dimensional adaptive kernel method;  the
mean biweight velocity $\overline{V}$; the projected radius $R$; 
and the robust velocity dispersion
$\sigma$ with the respective bootstrap errors computed by using the
ROSTAT routines by Beers, Flynn, \& Gebhardt (1990).  The structures
we found in clusters are highlighted in Figures~1 and~2. 

The centers of the two clumps are
only 0.1 and 0.06 \h away , in projection, for A1689 and A2218,
respectively, so that, considering the possible errors on the center
estimate, the clumps can be aligned along the line of sight.
Therefore A1689 and A2218 clusters could be cases
of {\em head-on} clumps merging.

It is not clear whether the redshift separation between the two clumps
comes from a difference in distance or rather it is due to a relative
motion by analyzing the magnitudes of galaxies.  The magnitudes of
galaxies belonging to the two clumps in A1689, S1 and S3, and A2218,
MS1 and MS2, are not different both according to the
Kolmogorov-Smirnov test (e.g.  Ledermann 1982, differing at 14\% and 20\% c.l.,
respectively) and the Median-test (Siegel 1956, differing at 60\% c.l.).  But,
if the mean velocity difference between clumps were simply due to
Hubble flow, the expected difference in magnitude would be so small
($\sim$ 0.1 mag) to be comparable with errors associated to the mean
values of magnitudes of substructure members.

\section{OPTICAL MASS ESTIMATES}

When the cluster appears far from dynamical equilibrium, showing two
or more clumps, it is more appropriate to compute its mass by adding
the masses of single clumps supposed to be virialized (e.g. Mohr,
Geller, \& Wegner 1996). In particular, in the case of the {\em
head-on} clumps merging, the velocity dispersion and the cluster mass
are strongly overestimated (e.g. G97; see also Pinkney et al. 1996 and
Roettinger et al.  1996 for simulated clusters).

The {\em standard} virial theorem, applied to galaxies within the
virialization radius $R_{vir}$, estimates the cluster mass as
$\displaystyle M=3\pi/2 \cdot \sigma^2 R_V/G$, 
where $R_V$ is the projected virial
radius which depends on the projected distance $r_{ij}$ between any
pair of galaxies, i.e.  
$\displaystyle  R_V=N^2/ \ \Sigma_{i \ne j} \ r_{ij}^{-1}$ where
$N$ is the number of galaxies (Limber \&
Mathews 1960; here we use the luminosity-unweighted expression).  
The virial mass estimate does not require any
assumptions about the isotropy of galaxy orbits (at least in the
context of spherical systems) but is stricly valid only if the mass
distribution follows galaxy distribution (e.g. Merritt 1988).
Virialization radii were computed similarly as in Bird (1995), by
assuming $R_{vir} \propto \sigma$ and by scaling to Coma values (see
\S~5.4 of G97).

%%TAB2
\hspace{-4mm}
\begin{minipage}{9cm}
\renewcommand{\arraystretch}{1.2}
\renewcommand{\tabcolsep}{1.2mm}
\begin{center}
\vspace{-3mm}
TABLE 2\\
\vspace{2mm}
{\sc Mass Estimates for A1689\\}
\footnotesize
\vspace{2mm}
\begin{tabular}{lrrrrr}
%\tablecaption{Mass Estimates for A1689}
\hline \hline
\multicolumn{1}{c}{Name}
&\multicolumn{1}{c}{$R_{vir}$}
&\multicolumn{1}{c}{$R_V$}
&\multicolumn{1}{c}{$M_{VT}$}
&\multicolumn{1}{c}{$M$}
&\multicolumn{1}{c}{$M$ range}
\\
&\multicolumn{1}{c}{(Mpc)}
&\multicolumn{1}{c}{(Mpc)}
&\multicolumn{1}{c}{($10^{14}$\m)}
&\multicolumn{1}{c}{($10^{14}$\m)}
&\multicolumn{1}{c}{($10^{14}$\m)}
\\
\hline
S1& 0.40&0.37& 0.59${\pm 0.17\phn}$& 0.46$_{-0.09\phn}^{+0.16\phn}$& (0.23-0.51)\\
S3& 0.49&0.45& 0.86${\pm 0.21\phn}$& 0.81$_{-0.11\phn}^{+0.21\phn}$& (0.40-0.91)\\
A1689&  &    & 1.45${\pm 0.38\phn}$& 1.27$_{-0.20\phn}^{+0.37\phn}$& (0.63-1.42)\\
\hline
\end{tabular}

\end{center}
\vspace{3mm}
\end{minipage}
\normalsize

In the case of A2218, the sampled region is much smaller than
$R_{vir}$. Thus, we have to assume the $\sigma$ computed in the
central region as the value of global velocity dispersion and to apply
a more complex procedure in the estimate of $R_V$.  We assume a
hydrostatic-isothermal surface density profile for the galaxy
distribution $\displaystyle \Sigma(r)=\Sigma_0 [1+(r/R_c)^2]^{-\alpha}$, where
$\Sigma_0$ is the central projected galaxy density and $R_c$ is the
core radius; this distribution corresponds to a volume-density density
$\rho(r)\propto r^{-2\alpha-1}$ for $r \gg R_c$.  We adopted a value
of $\alpha=0.8$, which gives a better fit to observational data than
the traditional King model (Girardi et al. 1995; see also Bahcall \&
Lubin 1994).  Then, adopting $R_c=0.17$ \h, i.e. the mean value
obtained for a sample of 90 nearby clusters, we computed $R_V$ at
$R_{vir}$ by means of the $R_V-R_c$ relation proposed by Girardi et
al. (1995, eq.~A8).

Note that this procedure, applied to A1689, gives a value of $1.27 \
10^{14}$ \msun, which agrees, within the errors, with 
the value of $1.45 \ 10^{14}$
\msun obtained by directly applying the virial theorem.\\

%%TAB3
\vspace{4mm}
\hspace{-4mm}
\begin{minipage}{9cm}
\renewcommand{\arraystretch}{1.2}
\renewcommand{\tabcolsep}{1.2mm}
\begin{center}
\vspace{-3mm}
TABLE 3\\
\vspace{2mm}
{\sc Mass Estimates for A2218\\}
\footnotesize
\vspace{2mm}
\begin{tabular}{lrrrr}
%\tablecaption{Mass Estimates for A2218}
\hline \hline
\multicolumn{1}{c}{Name}
&\multicolumn{1}{c}{$R_{vir}$}
&\multicolumn{1}{c}{$R_V$}
&\multicolumn{1}{c}{$M$}
&\multicolumn{1}{c}{$M$ range}
\\
&\multicolumn{1}{c}{(Mpc)}
&\multicolumn{1}{c}{(Mpc)}
&\multicolumn{1}{c}{($10^{14}$\m)}
&\multicolumn{1}{c}{($10^{14}$\m)}
\\
\hline
MS1& 0.81&0.50& 3.52$_{-1.29\phn}^{+3.11\phn}$& (1.71-4.13)\\
MS2& 0.98&0.48& 5.83$_{-1.01\phn}^{+1.89\phn}$& (2.86-6.99)\\
A2218&   &    & 9.35$_{-2.30\phn}^{+5.00\phn}$& (4.57-11.12)\\
\hline
\end{tabular}

\end{center}
\vspace{3mm}
\end{minipage}
\normalsize

In order to consider the possibility that the galaxy
 distribution of A2218
is characterized by
 a different core radius, we compute masses by considering two
boundary values for $R_c$: 0.01 \h and 0.5 \h.  Moreover, we compute
the same values also for A1689 to take into account the possibility
that the dark matter distribution differs from the galaxy distribution.  In
particular, small radii are suggested to be characteristic of dark
matter distributions (e.g. Durret et al. 1994).

In Tables~2 and ~3, for each clump of A1689 and A2218, 
 we list $R_{vir}$, $R_V$, the virial mass
$M$, as obtained using galaxy distribution,
 with the errors induced by $\sigma$, and the mass range 
corresponding to the above-mentioned range in $R_c$ (0.01-0.50 \h).
The mass values are  given for each system as a whole, too.
For A1689 we also list  the virial mass, $M_{VT}$, as obtained by directly
applying the virial theorem, with the respective jacknife error.

\section{MASS COMPARISON}

In order to compare our optical masses with the estimates derived from
X-ray and gravitational lensing analyses, we have to compute the
(projected) masses within the respective radii of comparison $R_{cp}$.
For each clump, we use the galaxy distribution adopted above (see
\S~3), constrained by the value of the mass at the virialization
radius. The large uncertainties adopted in the value of $R_c$ give
rise to large mass errors, in particular when small radii are
considered.

We computed the masses of the cluster as a whole by assuming that the clumps
are perfectly aligned along the line of sight; 
deviations from perfect alignment could lead us 
to overestimate the masses.\\

%%TAB4

\vspace{4mm}
\hspace{-4mm}
\begin{minipage}{9cm}
\renewcommand{\arraystretch}{1.2}
\renewcommand{\tabcolsep}{1.2mm}
\begin{center}
\vspace{-3mm}
TABLE 4\\
\vspace{2mm}
{\sc Mass Comparison\\}
\footnotesize
\vspace{2mm}
\begin{tabular}{lllllr}
%\tablecaption{Mass Comparison}
\hline \hline
\multicolumn{1}{c}{Name}
&\multicolumn{1}{c}{$R_{cp}$}
&\multicolumn{1}{c}{$M_{opt}$}
&\multicolumn{1}{c}{$M_X$}
&\multicolumn{1}{c}{$M_{gl}$}
&\multicolumn{1}{c}{Ref.}
\\
&\multicolumn{1}{c}{(Mpc)}
&\multicolumn{1}{c}{($10^{14}$\m)}
&\multicolumn{1}{c}{($10^{14}$\m)}
&\multicolumn{1}{c}{($10^{14}$\m)}
&
\\
\hline
A1689& 0.095 &0.14(0.08-0.26)&1.14$_{-0.75\phn}^{+2.0\phn}$ &1.8   & 1,2\\
A1689& 0.74  &2.22(0.82-3.84)&7.75                          &   -   & 3\\
A1689& 1.5   &3.43(1.12-6.54)&24  $_{-9.5\phn}^{+22\phn}$   &27    & 1,2\\
A2218& 0.0425&0.12(0.04-0.72)&0.16$_{-0.08\phn}^{+0.18\phn}$&0.305 & 1,4\\
A2218& 0.4   &4.40(3.04-4.44)&2.6 $_{-1.6\phn}^{+1.6\phn}$  &$>$3.9
$_{-0.7\phn}^{+0.7\phn}$  & 5\\
\hline
\end{tabular}

\end{center}
{\footnotesize\parindent=3mm
REFERENCES. 1~Wu
\& Fang 1997; 2~Tyson \& Fischer 1995; 3~White \& Fabian 1995; 4~Kneib
et al. 1995; 5~Squires et al. 1996.\\
\parindent=3mm
NOTES: All the masses, except for those in the second line,
refer to projected values.}
\vspace{3mm}
\end{minipage}
\normalsize

In Table~4 we compare our optical mass estimates, $M_{opt}$, with the
X-ray and gravitational lensing masses, $M_X$ and $M_{gl}$, as
presented in the literature within the respective projected radius
$R_{cp}$. The values of $M_{gl}$ refer to strong gravitational lensing
when they are given at small $R_{cp} (<0.1$ \h); otherwise they refer
to weak gravitational lensing.  As for our masses, we report also the
mass range obtained varying $R_c$ within the two boundary limits 
(0.01-0.50 \h); moreover, all
these mass values are affected by the errors due to velocity
dispersion, $\sim 25 \%$ and $\sim 40 \%$ for A1689 and A2218,
respectively (see Tables~2 and ~3).

In the case of A2218, our optical mass estimates agree, within the
errors, with those derived from X-ray and gravitational analyses, and
are significantly smaller in the case of A1689.

\section{DISCUSSION}

As discussed in this section,
 the most reliable interpretation of the results 
of our structure analysis  is that
A1689 and A2218 are cases of on-going merging along the line of sight. 

MB95 identified two clumps in the two-dimensional galaxy distribution of
clusters A1689 and A2218. However, they concluded that these clumps are too
dissimilar in size and not well enough aligned to produce a strong mass
overestimate by means of gravitational lensing analysis. 
The structures we detect do not exactly
correspond to the structures seen in the two-dimensional maps, since they are
comparable in size and superimposed along the line of sight, but they are
rather different in redshift. 
The presence of
on-going merging in cluster A2218 is also 
supported by the X-ray map and CCD images
in the study of Kneib et al. (1995, see also Squires et al. 1996)
and by its X-ray elongated shape (Wang \& Ulmer 1997).

The immediate consequence of our structure analysis is that our
optical mass estimates are  lower than the WF97 ones, which are
based on very high velocity dispersion, i.e. 1989 \ks for A1689
(Teague, Carter, \& Gray 1990) and 1370 \ks for A2218 (Le Borgne,
Pell\'o, \& Sanahuja 1992).

In the case of A1689, our optical mass estimates are significantly
smaller than those derived from X-ray and gravitational analyses.

The high value of X-ray temperature, 10.9 KeV, measured in A1689
(White \& Fabian 1995), which corresponds to the very high velocity
dispersion of 1350 \ks, could be due to the collision-heated gas, as
expected in numerical simulations (e.g. Schindler \& Mueller 1993;
Burns et al. 1995). There are some analogies with the case of A754, a
cluster with a X-ray temperature of 9 keV, whose two clumps have a
velocity dispersion of about 400-500 \ks and are colliding in the
plane of the sky (Zabludoff \& Zaritsky 1995, G97).  But, in our
cases, the merging is seen along the line of sight rather than in the
plane of the sky.  This fact leads to an apparent enhancement of the
observed velocity dispersion and can mask the typical features of
merging, which usually appear in two-dimensional galaxy distribution
and X-ray images (e.g. Schindler \& Mueller 1993). The high X-ray mass
estimate could be a direct consequence of the overestimate of X-ray
temperature. Note that Cirimele, Nesci, \& Trevese (1997), who used
the lower X-ray temperature of 10.1 KeV (David et al. 1993) found
a value for the X-ray mass of 5.9 \msun within 0.75 \h, which is
smaller than the estimate of White \& Fabian (1995) although
still larger than our optical estimate.

Any structure along the line of sight is particularly relevant
for masses derived from arcs, since the superposition of a modest mass
group can increase the surface mass density up to the critical value.
Since possible contribution could come
from all the structures between the source and A1689, we also
considered the whole cluster field.
A1689 appears well aligned along the line of sight with other
structures. Three foreground groups (with $\sigma \sim 500$ \ks) are
obvious and were already pointed out (Teague et al. 1990,
fig. 4). This fact suggests the presence of a large structure filament
well aligned along the line of sight. Indeed, our analysis of velocity
distribution (see \S~2 and Figures~1 and~2) detects, at a low significance
level, two other background structures roughly at the limit of the
redshift sample.  When assuming the reliability of these two
structures (with $\sigma \sim 750$ \ks), we can obtain a rough
estimate of their mass. Their contribution to the projected mass
within the lens-arc radius, i.e. $0.6 \cdot 10^{14}$ \msun or $1.5
\cdot 10^{14}$ \msun in the extreme case, can allow us to explain the
value of the lens-arc estimate.  
In this framework, also the sparse structure S2 can contribute a little, 
even if its contribution to the cluster mass is negligible. 
Only a larger and deeper redshift
sample could allow us to draw firm conclusions, especially with regard
to the presence of a large-scale structure filament.

In the case of A2218, Squires et al. (1996), whose X-ray mass agrees
with ours within errors, estimated an X-ray temperature of 3-5 $keV$,
which corresponds to $\sigma \sim$ 700-900 \ks.  These authors pointed out
that their estimate of temperature is smaller than that of ASCA and
GINGA, and suggested the presence of gas components with different
temperatures.  If this is the case, the higher temperature could
describe the gas heated by the collision between the merging clumps,
and the colder temperature would refer to the gas already roughly in
dynamical equilibrium.  In this scenario, the observed galaxy clumps,
which are more stable than the gas and can also survive the first
encounter of the clumps (McGlynn \& Fabian 1984), still preserve the traces of
the parent clumps.

As for the mass computed within the lens-arc, the large errors do not
allow us to draw firm conclusions about the discrepancy between lens
and X-ray masses, although we suspect that, for the aforesaid reasons,
the X-ray mass may not be reliable in the very central region.  Note
that the mass based on weak gravitational lensing at a larger radius
(Squires et al. 1996) is in good agreement with our mass estimate.

Also for the third cluster reported by MB95, A2163, the data analysis
shows evidence of recent merging (Elbaz, Arnaud, \& B\"ohringer 1995).
Another case of mass discrepancy, the A370 cluster (Wu 1994), shows a
bimodal structure (Kneib et al. 1993) and CL~0500-24 shows the
presence of two peaks in the velocity distribution (Infante et
al. 1994).  The presence of so many likely merging clusters at
moderate redshift is remarkable when compared with the percentage of
nearby clusters which show two or more peaks in the velocity
distributions (about 10\%, Fadda et al. 1996) or the presence of
strong substructures (about 14\%, G97). Indeed, there is evidence that
the probability for a cluster to form long arcs is significantly
larger if the cluster is substructured (Bartelmann, Steinmetz, \&
Weiss 1995; Bartelmann \& Steinmetz 1996), in particular if the
cluster is elongated along the line of sight (Miralda-Escud\'e 1993).
Note, however, that a good agreement of the mass determinations from
the strong gravitational lensing methods is found for the lowest
redshift cluster (z$\sim 0.10$) PKS~0745-191, which appears to be a
regular and relaxed cluster (Allen, Fabian, \& Kneib 1996).  Indeed,
only better statistics would make it possible to know whether clusters
which show long arcs are very atypical or whether distant clusters are
more substructured than nearby ones. This question is of considerable
importance for cosmology, since strong evidence of a
substructure-distance correlation could significantly constrain
theories of large-scale structure formation (Richstone, Loeb, \&
Turner 1992).

\section{SUMMARY AND CONCLUSIONS}

We summarize here our principal results:

i)~By applying a recent development of the method of wavelet analysis,
which uses the complete information obtained from optical data,
i.e. galaxy positions and redshifts, we find that both clusters A1689
and A2218 show the presence of structures superimposed along the line
of sight, suggesting that these clusters are cases of {\em head-on}
merging.

ii)~Our optical virial masses, estimated by
adding the masses of single clumps, are lower than the estimates of WF97,
who did not take into account the internal structure of clusters.

iii)~When comparing our optical masses with those derived from X-ray
and gravitational lensing analyses, we find a reasonable agreement in
the case of A2218, but we estimate a lower optical mass in the case of
A1689.

iv)~We suggest that the high X-ray mass of A1689 is due to the overestimated
X-ray temperature, which could be due to gas collision
phenomena, and we can explain the high projected gravitational lensing
mass only by adding the masses of two background galaxy
systems.

We stress that optical studies of clusters should be a default
complement of X-ray and gravitational lensing analyses.  However, only
optical analyses which consider the internal cluster structure can
allow one to draw reliable conclusions.  In particular, high values of
velocity dispersion for some clusters are probably due to the presence
of substantial subclustering along the line of sight.  In these cases,
the use of galaxy redshifts makes optical analysis a very suitable
method, while the standard X-ray and gravitational lensing analyses
suffer from strong deviation of gas hydrostatic equilibrium and from
the very atypical lens geometry, respectively.

For the clusters we analyze there is no evidence that present cluster
X-ray masses are underestimated; on the contrary, the mass of
A1689 appears overestimated.
If we generalize our results, we are still far from solving the 
so-called baryon crisis in clusters of galaxies by increasing
the estimates of cluster masses.

Moreover, there is increasing evidence that well-analyzed clusters
which have long gravitational arcs show the presence of a strong
substructure. This fact suggests that they are not the best targets
for making estimates of mass and baryon fraction, but rather it opens new
questions.  One needs to compare distant clusters selected on the basis
of lens-arc presence to a sample selected only on  X-ray and/or optical basis
(e.g., the CNOC redshift survey of distant clusters by Carlberg et
al. 1994) in order to understand whether they are unbiased examples of
distant clusters and to draw conclusions about cluster evolution.

\acknowledgments

We thank Daniel Gerbal for useful discussions  and the  anonymous
referee for his suggestions, which have allowed us to
improve the presentation of the paper. 
This work has been partially supported by the Italian Ministry of
University, Scientific Technological Research (MURST), by the 
Italian Space Agency (ASI), and by the Italian Research Council
(CNR-GNA).

\end{multicols}


\begin{references}

\footnotesize

\reference{} Allen, S. W., Fabian, A. C., \& Kneib, J. P. 1996, \mnras,
279, 615


\reference{} Bahcall, N. A., \& Lubin, L. M. 1994, \apj, 426, 513 


\reference{} Bartelmann, M. 1995, \aap, 295, 565 

\reference{} Bartelmann, M., \& Steinmetz 1996, \mnras, 283, 431

\reference{} Bartelmann, M., Steinmetz, \& Weiss 1995, \aap, 297, 1


\reference{} Beers, T. C., Flynn, K., \& Gebhardt, K. 1990, \aj, 100, 32

\reference{} Bird, C. M., 1995, \apj, 445, L81

\reference{} Burns, J. O., Roettiger, K., Pinkney, J., Perley, R. A.,
Owen, F. N., \& Voges, W. 1995, \apj, 446, 583

\reference{} Carlberg, R. G., Yee, H. K.  C., Ellingson, E., \&
 al. 1994, \jrasc, 88, 39

\reference{} Cirimele, G., Nesci, R., \& Trevese, D. 1997, \apj, 475, 11

\reference{} David, L. P., Slyz, A., Jones, C., Forman, W., Vrtilek,
S. D., \& Arnaud, K. A. 1993, \apj, 412, 479

\reference{} Durret, F., Gerbal, D., Lachieze-Rey, M., Lima-Neto, G,
\& Sadat, R. 1994, \aap, 287, 733

\reference{} Elbaz, D., Arnaud, M., \& B\"ohringer, H. 1995, \aap, 293, 337

\reference{} Escalera, E., \& Mazure, A. 1992, \apj, 388, 23.

\reference{} Evrard, A. E., Metzler, C. A., \& Navarro, J. F. 
1996, \apj, 469, 494

\reference{} Fadda, D., Girardi, M., Giuricin, G., Mardirossian, F.,
Mezzetti, M., \& Biviano, A., 1996, \apj, 473 670


\reference{} Girardi, M., Escalera, E., Fadda, D., Giuricin, G., Mardirossian, 
F., \& Mezzetti, M., 1997, \apj, 482, 41 [G97]

\reference{} Girardi, M., Fadda, D., Giuricin, G., Mardirossian, F.,
Mezzetti, M., \& Biviano, A., 1996, \apj, 457, 61


\reference{} Girardi, M., Giuricin, G., Mardirossian, 
Mezzetti, M., \& Biviano, A. 1995, \apj, 438, 527

\reference{} Gudehus, D. H., \& Hegyi, D. J. 1991, \aj, 101, 18

\reference{} Infante, L., Fouqu\'e, P., Hertling, G., Way, M.J., Giraud, E.,
\& Quintana, H. 1994, \aap, 289, 381

\reference{} Kneib, J. P., Mellier, Y., Fort, B., \& Mathez, G. 1993,
\aap, 273, 367

\reference{} Kneib, J. P., Mellier, Y., Pell\'o, R., Miralda-Escud\'e,
 J., Le Borgn\'e, J.-F., B\"ohringer, H. \& Picat, J.-P. 1995, \aap, 303, 27

\reference{} Le Borgne, J.-F., Pell\'o, R., \& Sanahuja, B. 1992,
\aaps 95,87

\reference{}Ledermann, W. 1982, Handbook of Applicable Mathematics,
eds. W. Ledermann and S. Vajda, Vol.6 

\reference{} Limber, D. N., \& Mathews, W. G. 1960, \apj, 132, 286

\reference{} McGlynn, T. A., \& Fabian, A. C. 1984, \mnras, 208, 709

\reference{} Merritt, D. 1988, in The Minnesota Lectures on Clusters of
Galaxies and Large-Scale Structures, ed. J. M. Dickey (ASP Conf. Ser., 5),
p. 175

\reference{} Miralda-Escud\'e, J. 1993, \apj, 403, 497

\reference{} Miralda-Escud\'e, J., \& Babul, A. 1995, \apj, 449, 18 [MB95]

\reference{} Mohr, J. J., Geller, M. J., \& Wegner, G.
 1996, \aj, 112, 1816

\reference{} Pinkney, J., Roettiger, K., Burns, J. O., \& Bird, C. A.
1996, \apjs, 104, 1 

\reference{} Pisani, A. 1993, \mnras, 265, 706

\reference{} Richstone, D., Loeb, A., \& Turner, E. 1992, \apj,
393, 477

\reference{} Roettiger, K., Burns, J. O., \& Loken, C. 1996, \apj, 473, 651


\reference{} Schindler, C. L. 1996, \aap, 305, 756

\reference{} Schindler, C. L. \& Mueller, E. 1993, \aap, 272, 137

\reference{} Serna, A., \& Gerbal, D. 1996, \aap, 309, 65

\reference{} Siegel S. in Non-paramteric Statistics for the 
Behavioral Sciences, 1956, cap.6, ed. McGraw-Hill 

\reference{} Squires, G., Kaiser, N., Babul, A., Fahlman, G.,
Woods, D. Neumann, D. M., \& B\"ohringer 1996, \apj, 461, 572

\reference{} Teague, P. F., Carter, D., \& Gray, P. M. 1990, \apjs,
72, 715

\reference{} Tyson, J. a., \& Fischer, P. 1995, \apjl, 446, 55

\reference{} Wang, Q. D.,\& Ulmer, M. P. 1997, \apjl, submitted,
preprint astro-ph/9702069

\reference{} White, D. A., \& Fabian, A. C. 1995, \mnras, 273, 72 

\reference{} White, S. D. M., Navarro, J. F., Evrard, A. E., \& Frenk, 
C. S. 1993, \nat, 366, 429 

\reference{} Wu, X. P. 1994, \apj, 436, L115

\reference{} Wu, X. P., \& Fang, L. Z., 1996, \apjl, 467, 45

\reference{} Wu, X. P., \& Fang, L. Z., 1997, \apj, 483, 62 [WF97]

\reference{} Zabludoff, A. I. \& Zaritsky, D., 1995, \apj, 447, L21.

\end{references}
\end{document}